\documentclass[a4paper, amsfonts, amssymb, amsmath, reprint, showkeys, nofootinbib, twoside, prb]{revtex4-1}
\usepackage[english]{babel}
\usepackage[utf8]{inputenc}
\usepackage[colorlinks,citecolor=red,urlcolor=blue,bookmarks=false,hypertexnames=true]{hyperref} 
\usepackage{url}

\usepackage{graphicx}
\usepackage[separate-uncertainty=true,multi-part-units=single,group-separator={,}]{siunitx}
\DeclareSIUnit[per-mode=symbol,per-symbol=p]{\ueV}{\micro\electronvolt}
\DeclareSIUnit[per-mode=symbol,per-symbol=p]{\um}{\micro\meter}

\usepackage{flushend}

\usepackage{xspace}
\newcommand{\cuprous}{Cu$_2$O\xspace}

\begin{document}
\title{Quantum confined Rydberg excitons in \cuprous nanoparticles}

\author{Konstantinos Orfanakis}
\author{Sai Kiran Rajendran}
\author{Hamid Ohadi}
\email{ho35@st-andrews.ac.uk}
\affiliation{SUPA, School of Physics and Astronomy, University of St Andrews, St Andrews, KY16 9SS, United Kingdom}
\author{Sylwia Zielińska-Raczyńska}
\author{Gerard Czajkowski}
\author{Karol Karpiński}
\author{David Ziemkiewicz}
\affiliation{UTP University of Science and Technology, Bydgoszcz, Poland}

\date{\today} 

\begin{abstract}
    The quantum confinement of Rydberg excitons is an important step towards exploiting their large nonlinearities for quantum applications. We observe Rydberg excitons in natural nanoparticles of \cuprous. We resolve up to the principal quantum number $n=12$ in a bulk \cuprous crystal and up to $n=6$ in nanoparticles extracted from the same crystal. The exciton transitions in nanoparticles are broadened and their oscillator strengths decrease as $\propto n^{-4}$ compared to those in the bulk (decreasing as $\propto n^{-3}$). We explain our results by including the effect of quantum confinement of exciton states in the nanoparticles. Our results provide an understanding of the physics of Cu$_2$O Rydberg excitons in confined dimensions.
\end{abstract}
\maketitle
\section{Introduction}
Solid-state quantum systems provide unprecedented capabilities for the 
realization of novel devices owing to their robustness, miniaturization capability and scalability~\cite{keyes_miniaturization_1988, awschalom_quantum_2013, devoret_superconducting_2013, loredo_scalable_2016,mak_photonics_2016, gonzalez-zalba_solid-state_2018}. The operation of such devices requires
developing means to efficiently produce, control and detect strongly 
interacting particles. Excitons, elementary excitations in semiconductors consisting
of a Coulomb-bound pair of an electron and a hole, are considered major candidates 
towards this direction. An exciton represents a solid-state analog of the hydrogen atom 
and hence excited states can be observed as a hydrogen-like discrete series 
at energies $R_y/n^2$ below the bandgap, with $R_y$ the Rydberg energy and $n$ the principal 
quantum number~\cite{fox_optical_2011}. Excitons in cuprous oxide (\cuprous) were observed as early as 1952~\cite{Hayashi_Hydrogen_1952}, and their various physical properties have been studied since then~\cite{Meyer_book_2013}. However only energy levels up to $n=8$ in \cuprous were observed for many decades until the Rydberg spectrum was extended to $n=12$ in 1996~\cite{matsumoto_revived_1996}.

Among semiconductors, \cuprous has the advantage of a large Rydberg energy which allows access to much higher excited states. In a recent high-resolution laser absorption study~\cite{kazimierczuk_giant_2014}, it was shown that 
\cuprous  hosts Rydberg excitons up to
$n=25$. This demonstration opened the portal to the field 
of giant Rydberg excitons in solid-state~\cite{thewes_observation_2015, grunwald_signatures_2016, heckotter_scaling_2017, heckotter_high-resolution_2017, takahata_photoluminescence_2018, kruger_waveguides_2018, lynch_giant_2018, mund_second_2019,zielinska-raczynskaNonlinearOpticalProperties2019}, in close analogy with their highly excited counterparts in atomic Physics~\cite{jones_special_2017}. Owing to their giant microscopic dimensions (up to \SI{\sim 1}{\um}) leading to the onset of exciton blockade, Rydberg excitons  in \cuprous exhibit enhanced optical nonlinearities at much smaller densities compared with other traditional semiconductors~\cite{kazimierczuk_giant_2014, walther_giant_2018}. 
These nonlinearities can be harnessed by quantum confinement of the excitons in semiconductor 
low-dimensional structures such as quantum wells and quantum dots~\cite{konzelmann_quantum_2020}.
Studying these excitons in confined dimensions is a crucial step towards harnessing these 
nonlinearities for applications. Nanoparticles~\cite{smith_semiconductor_2010} are naturally
a suitable system for quantum confinement and an interesting platform for realising quantum 
technologies with Rydberg excitons, potentially underpinning future devices such as 
single-photon sources\cite{khazali_single-photon_2017} and single-photon 
switches\cite{baur_single-photon_2014}.

In this work, we report the observation of Rydberg excitons in nanoparticles of
\cuprous. We resolve the yellow $p$-exciton states in natural nanoparticles up to $n=6$, while we observe up to $n=12$ in the bulk crystal that the particles were extracted from. We show that the reduction of the crystal structure from bulk
to nanoparticles leads to broadened linewidths and an apparent reduction of the oscillator strength of
the excitonic peaks compared to that in the bulk. We describe both effects by the quantum confinement of the Rydberg states and the size distribution of our nanoparticles.  Our results are the first demonstration of the effect of quantum confinement in linewidth and oscillator strength of Rydberg excitons.

\section{Experimental Methods}

\begin{figure}
  \includegraphics[width=1\linewidth]{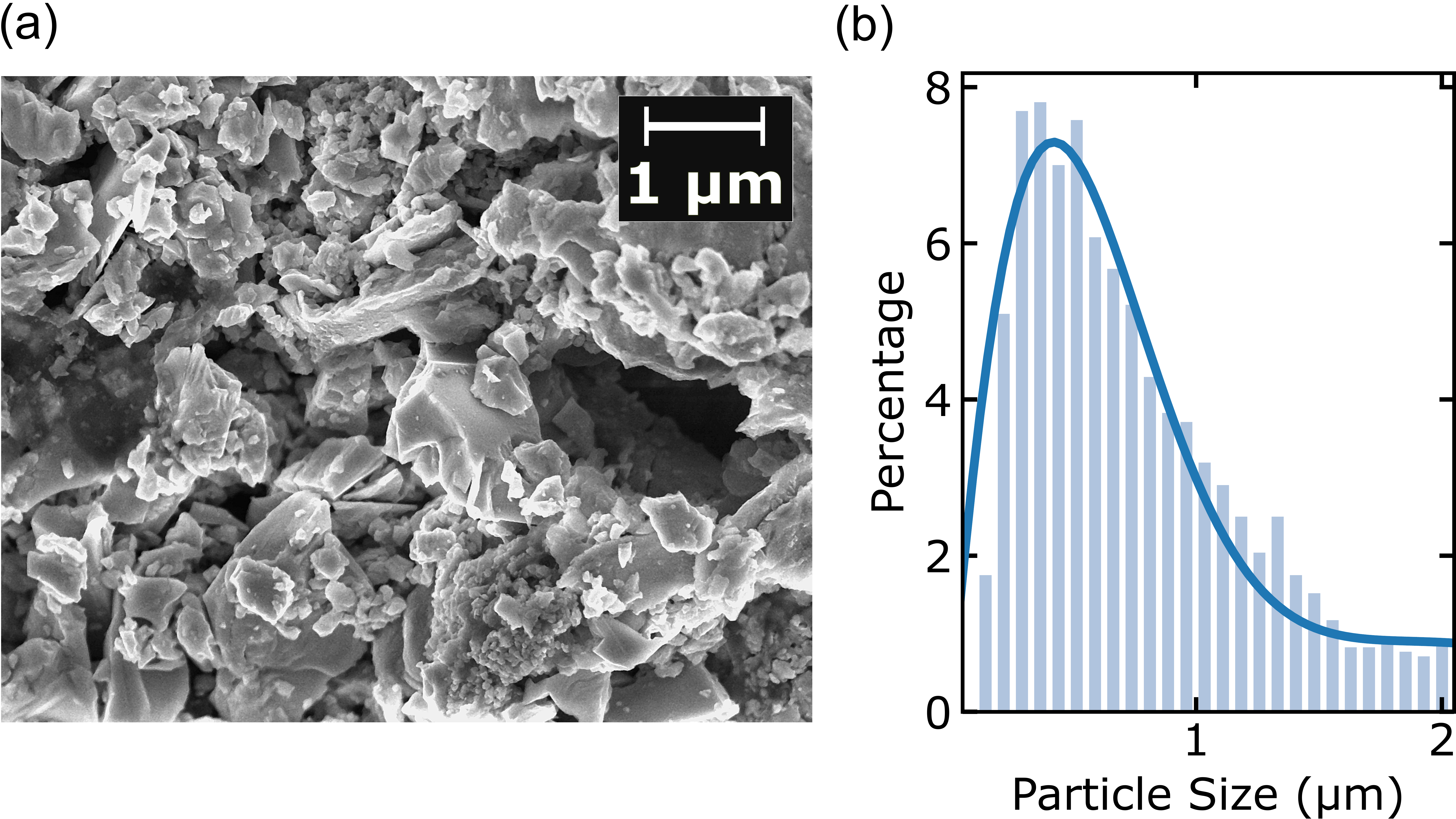}
  \caption{(a) SEM image of a cluster of particles. (b) Particle size distribution. The fitted distribution is depicted as a blue solid line. The mean diameter is calculated as $\simeq 1.1$ $\mu$m, while the median as $\simeq  0.7$ $\mu$m.}
  \label{fig:SEM}
\end{figure}

\cuprous nanoparticles were extracted from the residual powder from polishing a natural crystal mined in Tsumeb, Namibia. After forming a suspension of \cuprous powder in water, a droplet of this suspension is deposited onto a CaF$_2$ substrate. Scanning electron microscopy (SEM) reveals that \cuprous particles tend to aggregate
and form a thin layer due to Van der Waals forces as the particles 
redistribute during water evaporation [Fig.~\ref{fig:SEM}(a)]. The particles are of various sizes and shapes with a average and median diameter size of $\simeq 1.1$ and $\simeq  0.7$ nm [Fig.~\ref{fig:SEM}(b)]. We refer to these particles as natural nanoparticles (NNPs). Nanoparticles are compared to a thin slab of natural bulk crystal that is cut and
mechanically polished down to a thickness of \SI{~\sim 60}{\um}.  

We perform broadband transmission spectroscopy (see Supplementary Information 1 for setup). The excitation source 
is a green-yellow light-emitting diode (LED) with
a center wavelength of \SI{554}{\nm} (Thorlabs MINTF4). The resulting signal is 
collected, dispersed and analyzed in a spectrometer (Andor Shamrock 750) coupled to a
CCD camera. For transmission spectroscopy, an objective lens (20$\times$
Mitutoyo Plan Apo, NA = 0.42) focuses the excitation light to a spot on our sample ($\sim$100 $\mu$m diameter)
and a second objective lens (same NA as excitation) collimates the 
transmitted light before it enters the spectrometer. The sample is maintained at
\SI{4}{\K} using a liquid-helium flow cryostat. We use CaF$_2$ substrates because of their 
excellent thermal conductivity at cryogenic temperatures and transparency in the 
visible spectrum.

\section{Experimental results}
\begin{figure}
  \includegraphics[width=0.9\linewidth]{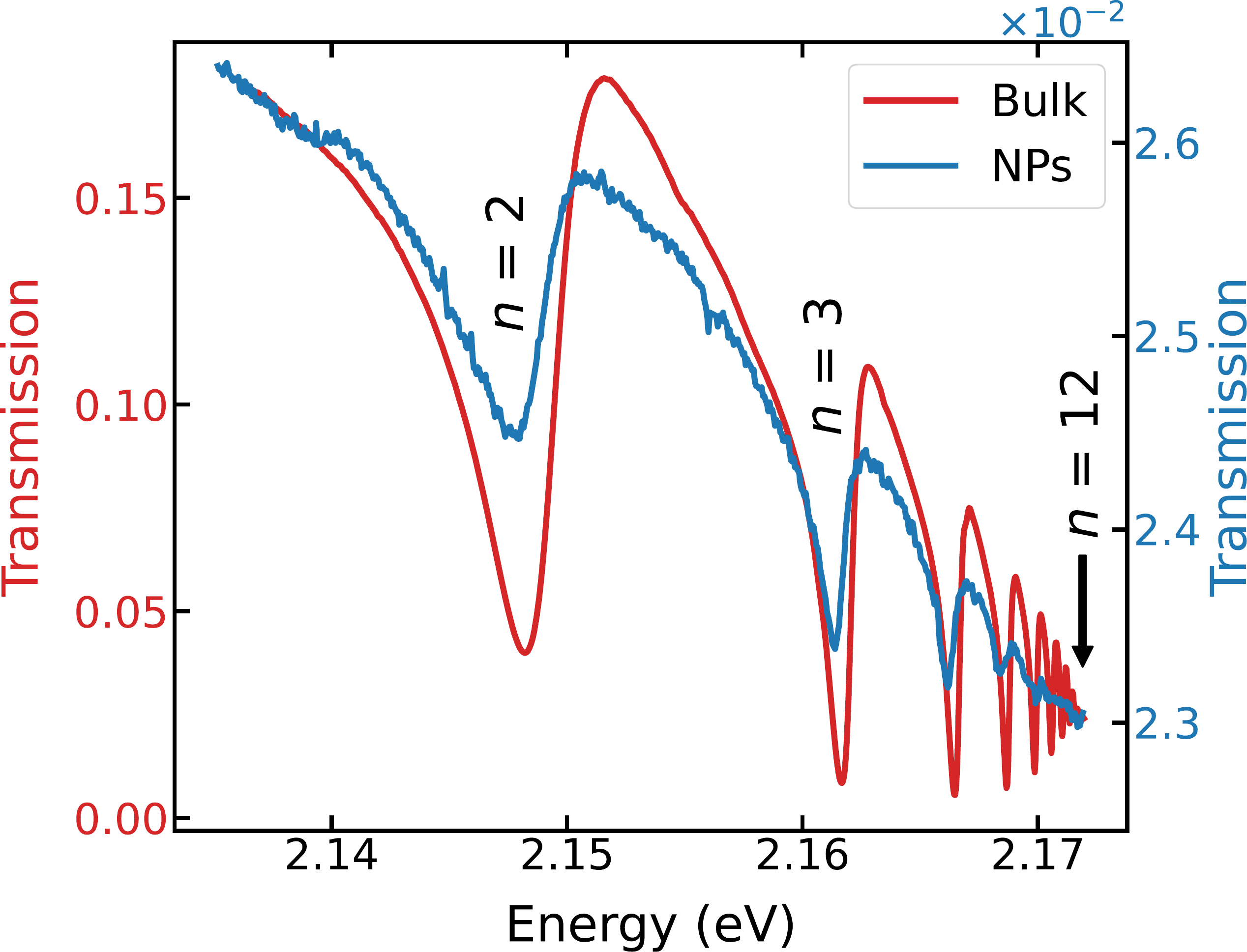}
  \caption{Transmission spectrum for a natural crystal of \cuprous at \SI{4}{\K}, where excitons up to $n = 12$ are resolved (red line) and for natural nanoparticles (blue line).}
  \label{fig:SetupAndWholeSpectra}
\end{figure}

The transmission spectrum of the bulk crystal reveals a series of absorption lines corresponding to the excited states of Rydberg excitons in \cuprous [red line in Fig.~\ref{fig:SetupAndWholeSpectra}(b)]. These states are labelled by their principal quantum number, $n$.  We can reliably identify Rydberg states up to $n=12$ overlaid
on a continuous phonon background~\cite{baumeister_optical_1961, jolk_linear_1998}. The reduced number of exciton lines observed in our experiment compared to previous works ~\cite{kazimierczuk_giant_2014} is primarily attributed to the small diameter and broadband spectrum of our excitation source. The asymmetric lineshape stems from the Fano
interference between the discrete excitonic states and the absorption background 
originating from phonon-assisted absorption of the 1$s$ exciton~\cite{toyozawa_interband_1964}. By fitting the 
exciton resonance energies to the Rydberg formula, we extract the bandgap energy 
$E_g = \SI{2.173}{\eV}$ and the Rydberg energy $R_y = \SI{97.1}{\meV}$,  which 
agree with those reported in the literature~\cite{agekyan_fine_1974, matsumoto_revived_1996}.

The transmission spectra of the nanoparticles covering a large area ($\sim$100 $\mu$m diameter) on the substrate show clear exciton peaks [blue line in Fig.~\ref{fig:SetupAndWholeSpectra}(b)]. Exciton resonances are evident as discrete asymmetric dips on top of the phonon background. A noticeable difference between the spectra of 
clusters and that of the thinned bulk crystal is that, in the former, is the sharp decrease in the peak absorption of the excitons as $n$ increase, such that we can only
resolve resonances up to $n = 6$. 

\begin{figure}
    \centering
    \includegraphics[width=0.85\linewidth]{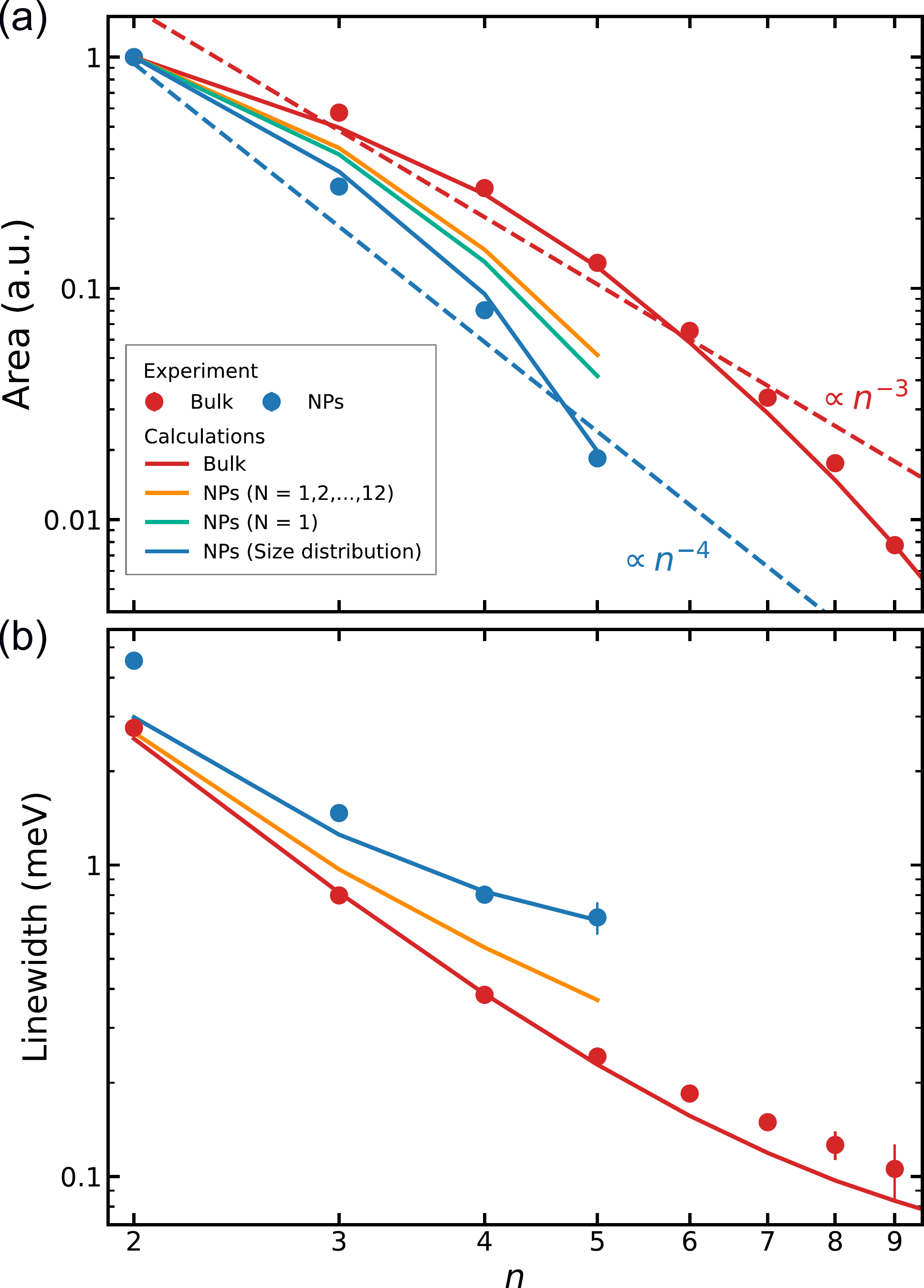}
    \caption{Peak areas (a) and linewidths (b) calculated from experimental (dots) and theoretical results (lines). Peak areas are normalized to the value of $n$ = 2 in each case.}
    \label{fig:Areas}
\end{figure}

Repeating the measurements on synthetic nanoparticles (SNPs) of comparable sizes can show if the reduced number of observed excitonic resonances in NNPs could be due to the polishing procedure. The transmission spectrum for SNPs, exhibits the same number of resonances as the one for NNPs (see Supplementary Information 2). Therefore, we rule out polishing as the reason for the observation of the reduced number of transitions. We note that resonances with the same $n$ are slightly blueshifted in the bulk crystal compared to those in nanoparticles due to a small temperature variation in each experiment (see Supplementary Fig. 3).

The comparison of oscillator strength (peak area) of the excitons in bulk and NPs shows (see Fig.~\ref{fig:Areas}a) that the relative peak area of excitons in nanoparticles of \cuprous decreases as $\propto n^{-4}$ compared to that in the bulk (decreasing as $\propto n^{-3}$, which is the theoretical dependence). Individual peaks were fitted with an asymmetric Lorentzian~\cite{kazimierczuk_giant_2014} to extract their linewidth. The linewidth of NPs is approximately double the linewidth of the bulk thin crystal for the first three observed resonances. The  broadening is more pronounced for $n = 5$ as the linewidth increases by nearly three folds for NPs (Fig.~\ref{fig:Areas}b).

Previously, the broadening of excitons in nanoparticles of CuCl (\SI{<15}{\nm} diameter) was found to be due to quantum confinement and size distribution~\cite{Ekimov_CuCl_1985, Ekimov_CuCl_1986}. Here, we show that quantum confinement results in the broadening of the transitions in a similar process, however, due to the narrow linewidth of the transitions the effect is visible for larger nanoparticle sizes.

\section{Theoretical description}
\begin{figure}
    \centering
    \includegraphics[width=\linewidth]{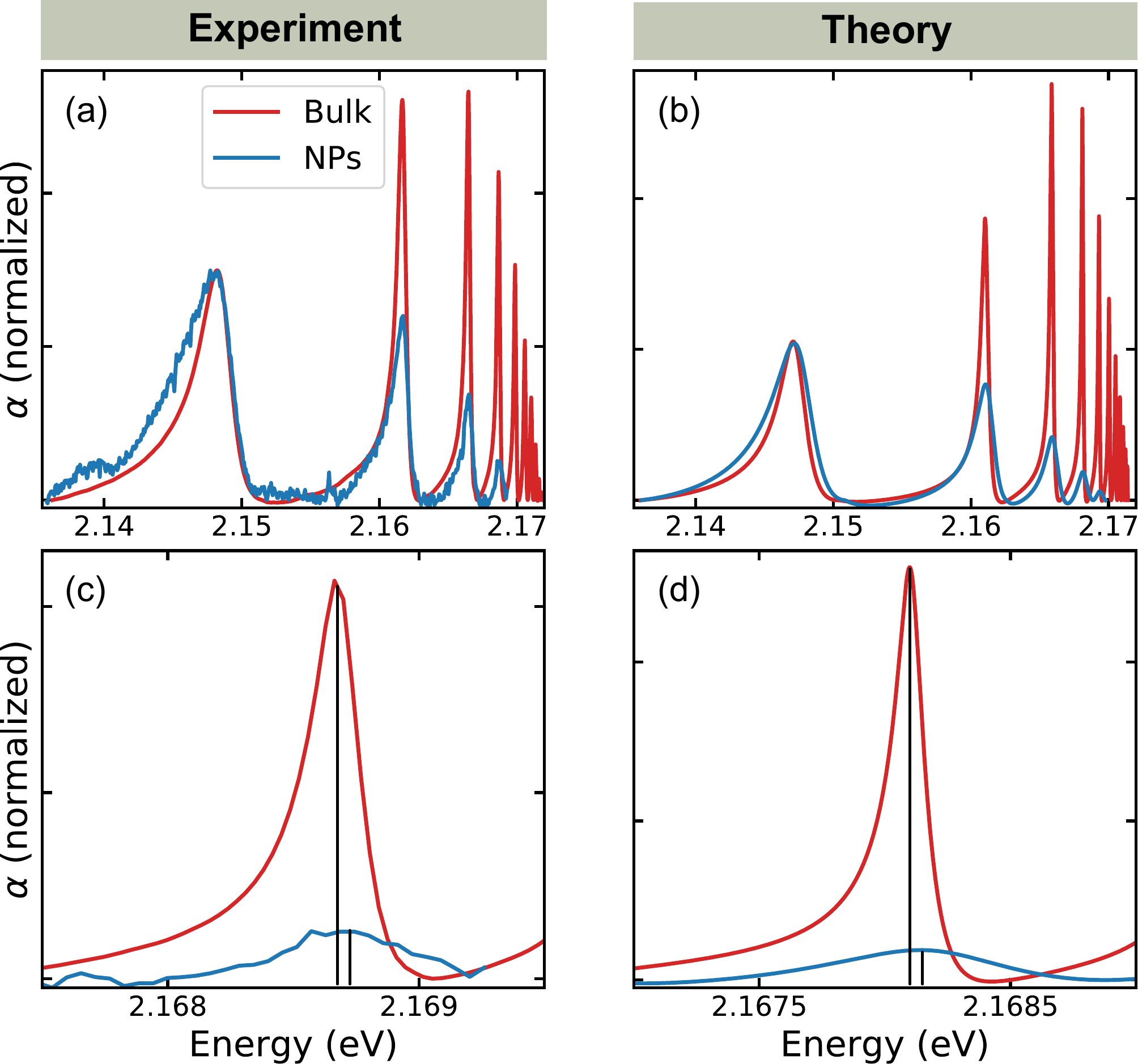}
    \caption{Experimental (a),(c) and theoretical (b),(d) absorption spectra of bulk \cuprous NPs in the full energy range and in the vicinity of $n=5$ state.}
    \label{fig:Comparison2}
\end{figure}

From the point of view of the classification of low-dimensional semiconductor structures, the considered nano-crystals are quantum dots (QDs). The SEM images (Fig. \ref{fig:SEM}) suggest that the best model shape is a spherical quantum dot, with a median diameter of $\sim$700~nm. 
Assuming that the interaction between dots is small enough, we may consider interband transitions in an isolated QD, and then average the results over a size distribution of QDs. As it has been shown previously\cite{Ziel_prb2020}, the optical properties of QDs can be studied with the real density matrix approach (RDMA). In this method, the optical response (absorption, reflection,
transmission) is obtained by solving a set of equations for the
excitonic amplitude $Y(\textbf{r}_e,\textbf{r}_h,t) $ and electric field vector $\textbf{E}(\textbf{R},t)$ of the wave propagating in the nanostructure. The quantities
$\textbf{r}_e,\textbf{r}_h$ are the coordinates of the electron
and the hole, and \textbf{R} is the center-of-mass coordinate of
the electron-hole pair. The basic equations of RDMA have the form
\cite{Ziel_prb2020}
\begin{equation}\label{Ylinear}
-i(\hbar \partial_t + {\mit\Gamma})
Y(\textbf{r}_e,\textbf{r}_h,t)+H_{eh}Y(\textbf{r}_e,\textbf{r}_h)
={\bf M}{\bf E},
\end{equation}
\noindent where $\mit\Gamma$ is a phenomenological damping
coefficient, ${\bf M}(\textbf{r})$ is a smeared-out transition
dipole density, $E_g$ is the fundamental gap, and
$\textbf{r}=\textbf{r}_e-\textbf{r}_h$ is the relative
electron-hole distance. The operator $H_{eh}$ stands for
 the two-band effective mass Hamiltonian, which includes the electron and hole kinetic energy, the
electron-hole interaction potential and the confinement
potentials. In consequence, the Hamiltonian $H_{eh}$ is given by
\begin{eqnarray} \label{hamilt}
&&H_{eh}=E_{g} +\frac{{\bf
p}_{h}^2}{2m_h} \nonumber\\
&&+\frac{{\bf p}_{e}^2}{2m_{e}}+V_{eh}(\textbf{r}_e,\textbf{r}_h)
+V_h(\textbf{r}_h)+V_e(\textbf{r}_e),
\end{eqnarray}
\noindent where the second and the third terms on the r.h.s. are
the electron and the hole kinetic energy operators with
appropriate effective masses, the  fourth term is the
electron-hole attraction, and the two last terms are the surface
confinement potentials for the electron and hole.
The total
polarization of the medium is related to the coherent amplitude by
\begin{equation}\label{Polar}
{\bf P}({\bf R})=2 \hbox{Re}\int d^3{r}\,{\bf M}({\bf r}) Y({\bf
R},{\bf r}).
\end{equation}
 This, in
turn, is used in Maxwell's field equation
\begin{equation}\label{Maxwell}
c^2\hbox{\boldmath$\nabla$}^2 {\bf E(R)} - \epsilon_b \ddot{\bf E}
= \frac{1}{\epsilon_0}{\bf \ddot{P}(R)},
\end{equation}
where $\epsilon_b$=7.5 is the QD material dielectric constant.
Equations (\ref{Ylinear})-(\ref{Maxwell}) form a system of coupled
integro-differential equations in 6-dimensional configuration
space $(\textbf{r}_e,\textbf{r}_h)$
The optical properties of spherical
QDs can be described by means of the exciton center-of-mass (COM)
quantization method\cite{Ziel_prb2020}. One assumes that the COM is confined within a
sphere of radius $R_{\scriptsize\rm{max}}$, which will give the
confinement states. Those states will overlap with the
3-dimensional exciton states. The effective QD susceptibility in
this limit is given by the formula (see supplementary information)
\begin{eqnarray}\label{main_eq}
&&\overline{\chi}_{QD}=\epsilon_b\sum\limits_{n=2}^{n_{max}}\sum\limits_{N=1}^{N_{max}}\frac{f_{n1}f_N\Delta^{(2)}_{LT}/R_y}{\left(E_{Tn}-E-
i{\mit\Gamma}\right)/R_y +\frac{\mu}{M_{tot}}\left(\frac{x_{\frac{3}{2},N}}{R}\right)^2},\nonumber\\
 &&f_N=\frac{1}{6\pi}\left|j_2(x_{\frac{3}{2},N})\right|^{-2}\,x_{\frac{3}{2},N}^2\left[\int\limits_0^1u^3du\,\prod\limits_{s=1}^\infty\left(1-\frac{x_{\frac{3}{2},N}^2}{x^2_{\frac{3}{2},s}}u^2\right)
 \right]^2,\nonumber\\
 &&R=\frac{R_{\scriptsize\rm{max}}}{a^*},\\
&&f_{n1}=\frac{32(n^2-1)}{3n^5}\left[\frac{nr_0(r_0+2a^*)}
{2r_0(r_0+na^*)}\right]^6,\nonumber
 \end{eqnarray}
where $j_N(x)$ are the spherical Bessel functions ($N=1,2,\ldots$),
$x_{N,s}$ are roots of the equation $j_N(x)=0$, $s=1,2,\ldots$, $a^*=1.1$ nm is the Rydberg radius and $E_{Tn}$ are energies of excitonic levels.

Fig.~\ref{fig:Comparison2} shows a comparison of experimental and calculated absorption spectra. 
The bulk crystal spectrum is also calculated from Eq. (\ref{main_eq}) by taking a very large NP radius $R>$\SI{10}{\um}. Due to the relatively large size of considered NPs and the corresponding confinement energy shifts on the order of $\sim$ \SI{10}{\ueV}, the direct observation of them can be challenging. To get a clearer picture of the results, the absorption coefficient obtained for bulk and nanoparticle systems has been normalized and the NP spectrum was shifted to obtain an exact match of $n=2$ lines, removing the influence of different experimental conditions such as temperature mentioned before. In such a case, we observe a difference of $\Delta E \sim$ \SI{25}{\ueV} between bulk and NP lines for $n=5$ state (vertical lines on Fig. \ref{fig:Comparison2} (c) and (d)), both in the experimental spectra and calculation results. This amounts to the difference between confinement energy of $n=5$ and $n=2$ state, the latter one being negligible. This result is a close match to the theoretical predictions of Konzelmann \emph{et al.}~\cite{konzelmann_quantum_2020}. 

Our calculations show an excellent match between the measured oscillator strength and the theoretical estimates when the size distribution of the nanoparticles in considered [blue line in Fig.~\ref{fig:Areas}(a)]. The calculation for single NP size ($R$=350 nm) fails to fully explain the oscillator strength reduction [orange line in Fig.~\ref{fig:Areas}(a)], further confirming that quantum dot size distribution is crucial to the understanding of the observed effects. Moreover, upper confinement states ($N>1$) contribute to $\sim$23\% of the peak area for $n=5$, with a smaller effect for $n<5$. The calculated linewidths match the experimental data for $n=3,4,5$, while slightly underestimating the linewidth of $n=2$ state [blue line in Fig.~\ref{fig:Areas}(b)]. Again, we find that it is crucial to take into account the size distribution of the NPs, further evidencing the effect of NP size.

\section{Discussion}
\begin{figure}
    \centering
    \includegraphics[width=0.8\linewidth]{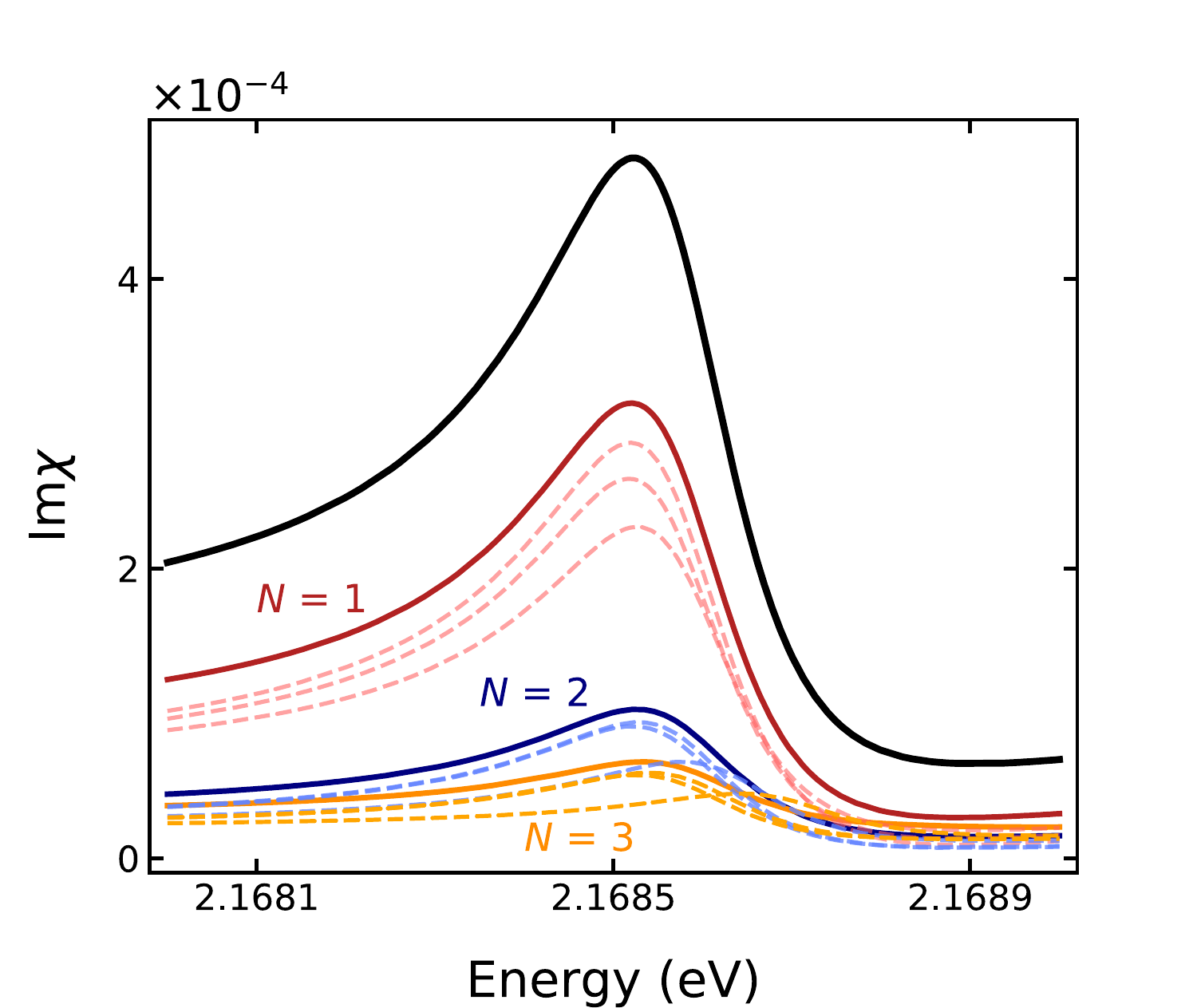}
    \caption{Imaginary part of susceptibility calculated for $n=5$ and various $N$ values and dot sizes. For each $N$, a sample of three individual dot sizes is shown (dashed lines) along with an average (solid lines). The normalized sum over all confinement states and dot sizes is depicted as a black solid line.}
    \label{fig:bdetail}
\end{figure}

Previous studies~\cite{kazimierczuk_giant_2014} have shown that the power-dependent optical bleaching (reduction of peak area) only becomes visible for $n=12$ for a pump intensity of $\sim$1 mW/mm$^2$. Moreover, the intensity required to see optical bleaching scales as $n^{10}$. Therefore, to notice any nonlinearities at $n=6$ one needs laser power on the order of 1 W/mm$^2$. The intensity of the illuminating light in our experiment ($\sim 1$ mW/mm$^2$) is sufficiently low that the nonlinear effects observed in bulk can be neglected for the observed excitonic states. We note that in nanoparticles, the blockade diameter can become comparable to particle size for low $n$ (see Supplementary Fig. S4). The intensities in our experiment, however, are 3 orders of magnitude weaker than that required to reach the blockade effect.

The reduction in oscillator strength due to damage to crystal structure can be ruled out since we observe this reduction in natural as well as synthetic NPs (see Supplementary Information 2). Strain on \cuprous crystals can cause a change in the absorption strength due to change from isotropic to anisotropic $np$ states~\cite{Agekyan_review_1977}. The substrates can exert stress on \cuprous crystals and affect the exciton transitions especially if the crystal is glued onto the cold finger or strongly sandwiched between two substrates. However, the nanoparticles here are held in place only by the van der Waals forces.

Electron-hole plasma generated due to the incident broadband light can affect the excitons. The collision of excitons with electron-hole plasma results in the reduction of exciton lifetime, broadening of the transitions as well as reducing the exciton oscillator strength. However, this effect was observed~\cite{Heckotter_plasma_2018} to be significant only for $n$ levels higher than $n = 10$.

The quantum confinement, however, is the dominant effect here.
Theoretically, for $n= 5$ excitons in \cuprous three dimensional confinement to \SI{700}{\nm} diameter, the lowest confinement state ($N$=1) would exhibit a blueshift of $\sim$ \SI{20}{\ueV}~\cite{konzelmann_quantum_2020}, which is an order of magnitude smaller than the linewidth of the $n=5$ transition. However, the states $N=2,3,4...$ provide a non-trivial contribution to the total area of the observed excitonic line, which is a measure of oscillator strength. This is shown on the Fig. \ref{fig:bdetail}, where $N=1,2,3$ states are marked by red, blue and green lines respectively. Since the energy shift of those states is proportional to $N$, it can reach values of over \SI{200}{\ueV} for $N>2$. Moreover, for the given dot radius $R$, the energy shift is approximately proportional to $R^{-2}$ and a relatively large energy shift on the order of \SI{0.5}{\meV} can be expected at the lower end of the obtained NP sizes. This is clearly seen in Fig. \ref{fig:bdetail}, where strongly blueshifted peaks corresponding to small NPs are visible. All these factors contribute to the shape of the total line (black curve) and result in an apparent reduction of the oscillator strength; every observed excitonic line is an overlap of the primary confinement state ($N=1$) and multiple blueshifted higher states which are too close to each other to discern them on the spectrum. The contribution of these states is twofold. 1- They make the transitions more symmetric. Since they have higher energy than the $N=1$ state, their effect is more pronounced on the right-hand side of the absorption peak; this is visible in Fig.~\ref{fig:Comparison2}(c) and (d) where the asymmetric lineshape of the bulk crystal is transformed into Gaussian shape. The effect is also visible for lower $n$ states [Fig. \ref{fig:Comparison2}(a) and (b)], but it is much less pronounced due to smaller energy shifts. 2- They significantly increase the background absorption. Since the oscillator strength of confinement states quickly decreases with $N$, their contribution affects mostly the base of the absorption line. The widened bases overlap, forming a strong absorptive background. This greatly reduces the area of the peak visible above that background resulting in an apparent reduction of the oscillator strengths.

In conclusion, we successfully observed Rydberg excitons in nanoparticles of \cuprous. Through optical spectroscopy, we showed that reducing the size of the system leads to a subsequent reduction in the oscillator strength and an apparent linewidth broadening. We explained our observations through the quantum confinement of the excitons in the nanoparticles. Our work paves the way for exploiting Cu$_2$O Rydberg excitons in the nanoscale for their large nonlinearities.

\section*{acknowledgements}
We acknowledge EPSRC Grant No. EP/S014403/1 and The Royal Society RGS\textbackslash R2\textbackslash 192174.  K.O. acknowledges EPSRC for PhD studentship support through grant no. EP/L015110/1. We thank Michael Huang, Matthew Jones, Stephen Lynch, Stefan Scheel and Mikhail M. Glazov for fruitful discussions.

%


\end{document}


\renewcommand{\theequation}{S\arabic{equation}}
\renewcommand{\thefigure}{S\arabic{figure}}
\setcounter{figure}{0}

\date{\vspace{-5ex}}
\title{Supplementary information: Quantum confined Rydberg excitons in Cu$_2$O nanoparticles}\maketitle
\begin{center}
Konstantinos~Orfanakis, Sai~Kiran~Rajendran, Hamid~Ohadi\footnote{ho35@st-andrews.ac.uk}\newline\\

SUPA, School of Physics and Astronomy, University of St Andrews, St Andrews, KY16 9SS, United Kingdom,\newline\\

Sylwia~Zieli\'{n}ska-Raczy\'{n}ska, Gerard~Czajkowski,  Karol~Karpi\'{n}ski, David~Ziemkiewicz\newline\\

UTP University of Science and Technology, Bydgoszcz, Poland\\
\end{center}

\section{Setup}

\begin{figure}[h!]
\centering
\includegraphics[width=0.8\linewidth]{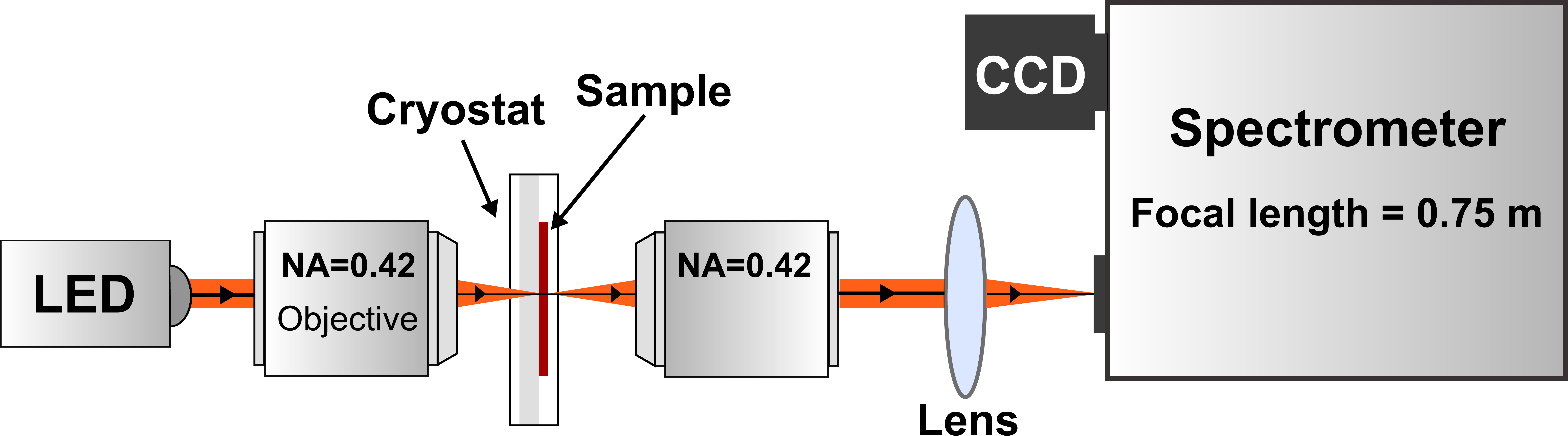}
\caption{Experimental setup for transmission spectroscopy. Broadband light is focused and collected using two objective lenses (NA = 0.42, and $f = 20$ mm).}
\end{figure}

\section{Synthetic nanoparticles}

\begin{figure}
\centering
\includegraphics[width=.5\linewidth]{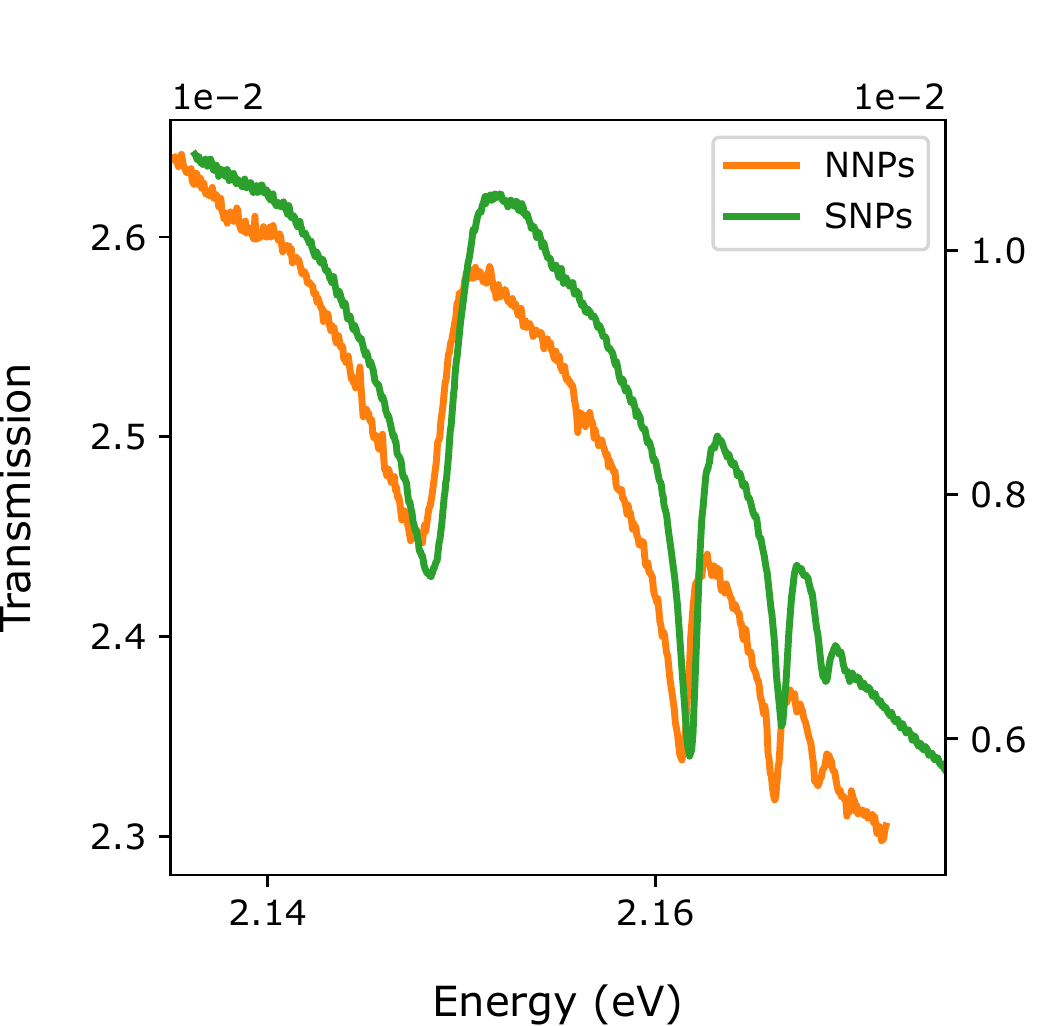}
\caption{Transmission spectrum for NNPs (green) and SNPs (orange). }
\label{sfig:SNP}
\end{figure}

We compare the commercial synthetic nanoparticles (Nanografi) with an average size distribution of 0.6~$\mu$m with our natural nanoparticles. We refer to these as synthetic nanoparticles (SNPs). Comparison of the two spectra shows a similar number of transitions, and broadening of the transitions (Fig.~\ref{sfig:SNP}).

\section{Temperature dependence}
\begin{figure}
\centering
\includegraphics[width=.7\linewidth]{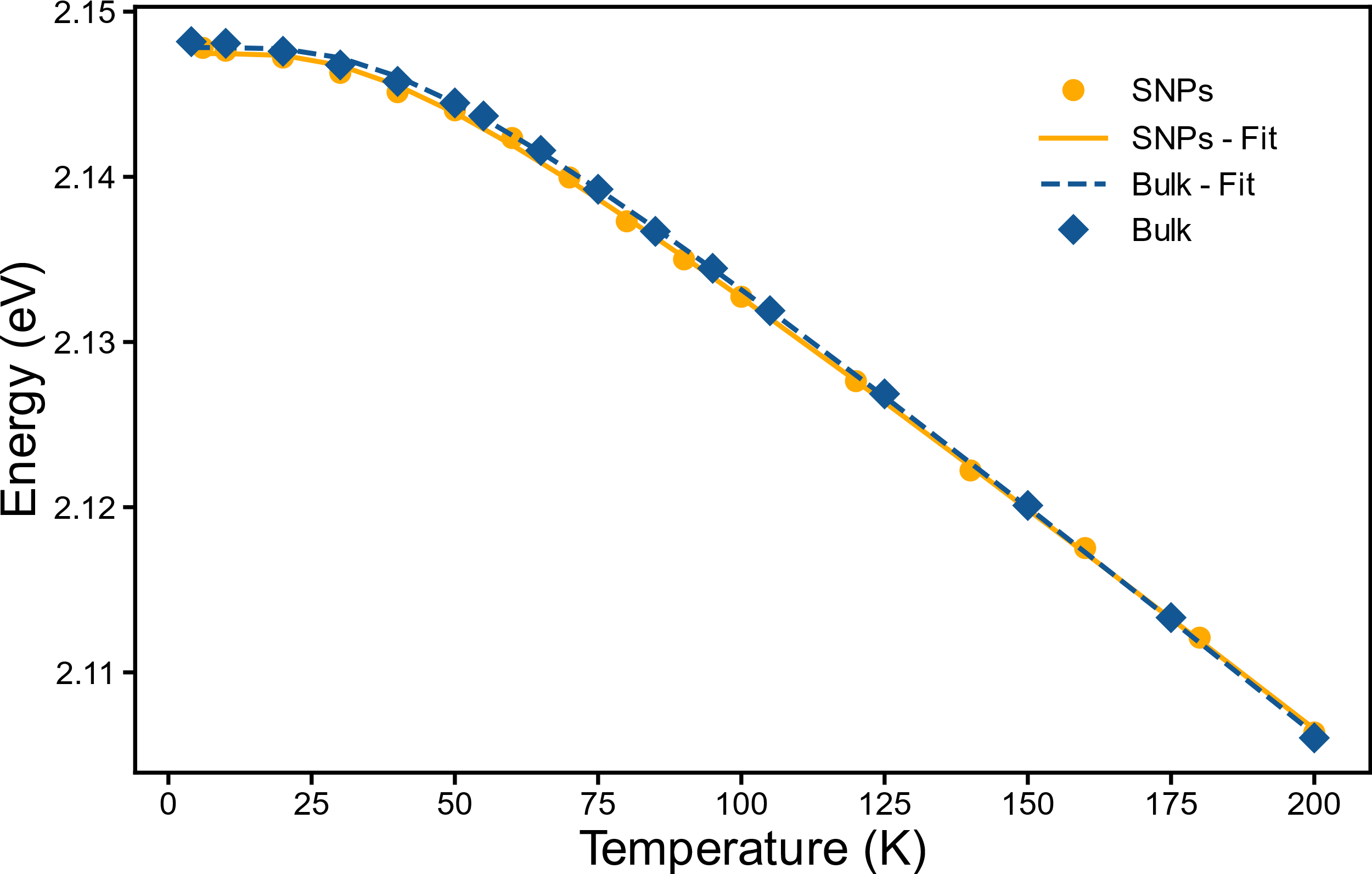}
\caption{Variation of the  exciton line with temperature for SNPs and bulk crystal of Cu$_2$O, shown as orange dots and blue diamonds, respectively. The corresponding fitted curve using Eq. S1 is shown as a line in each case.}
\label{sfig:Tdep}
\end{figure}

The temperature evolution of the exction transmission line for boths SNPs and bulk crystal of Cu2O is shown in Fig.~\ref{sfig:Tdep}. The temperature range spans from 4 to 200 K, the highest temperature where the exciton line can be resolved in our system. The increase of temperature results in a progressive red-shift in the exciton transition energy, a behaviour characteristic of semiconductors.
The variation of the bang-gap of semiconductor  can be described in terms of a hyperbolic cotangent relation [1]:
\begin{equation}
E_g(T)=E_g(0)-S\langle\hbar\omega\rangle\left[\coth(\hbar\omega/2k_BT)-1\right],
\label{eq:Tdep}
\end{equation}
where $E_g(0)$ is the band gap energy at 0 K, $S$ is a dimensionless constant describing the strength of the electron-phonon coupling, $k_B$ is the Boltzmann constant, and $\langle\hbar\omega\rangle$ is an average phonon energy. This equation can be used to describe the redshift of the $n=2$ line with parameters summarized in Table~\ref{tab:Tfits}.

\begin{table}
\centering
\begin{tabular}{|c|c|c|}
 \hline
 & NPs & Bulk crystal\\\hline
$E_g(0)$ & 2.1475 & 2.1478\\
S & 1.60 & 1.66\\
$\langle\hbar\omega\rangle$ (meV) & 9.8 & 10.5\\\hline 
\end{tabular}
\caption{Fitting parameters of temperature dependence of the exciton transitions in NPs and bulk using Eq.~\ref{eq:Tdep}.}
\label{tab:Tfits}
\end{table}

\section{Blockade volume}
\begin{figure}
\centering
\includegraphics[width=.7\linewidth]{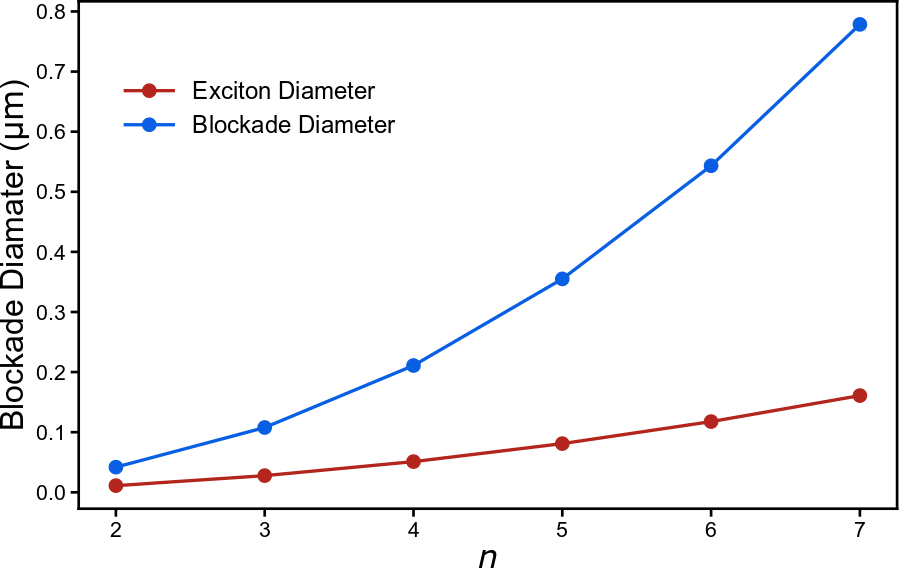}
\caption{Exciton (red) and Blockade (blue) diameter as a function of principal quantum number $n$. The solid line is used as a guide to the eye.}
\label{sfig:blockade}
\end{figure}

Estimation of the dipole blockade volume, hence blockade radius, was done using the following equation [2]:
\begin{equation}
V_{blockade}=3\cdot10^{-7}\mu \mathrm{m}^3n^7.
\end{equation}
For $n=6$, the first transition we cannot resolve in nanoparticles, the blockade diameter is $\sim$ 540 nm, i.e. approximately equal to the average diameter of our SNPs (Fig.~\ref{sfig:blockade}). For comparison, the average exciton diameter for $n=6$, is approximately 120 nm calculated from
\begin{equation}
d_n=a_b(3n^2-l(l+1)),
\end{equation}
where $a_b=1.11$~nm  is the Bohr radius for P-excitons [2].
\clearpage
\section{Calculation of Cu$_2$O quantum dot susceptibility}
The Hamiltonian (2) for larger nanocrystals, when the
relative electron-hole motion can be separated from the
center-of-mass motion, has the form [3]
\begin{eqnarray*}
&&H_{\scriptsize{eh}}=E_g+H_{\scriptsize{\textbf{r}}}+H_{\scriptsize{\textbf{R}}},\\
&&H_{\scriptsize{\textbf{r}}}=-\frac{\hbar^2}{2\mu}\hbox{\boldmath$\nabla$}^2_{\scriptsize\textbf{r}}+V(r),\\
&&H_R=-\frac{\hbar^2}{2M}\hbox{\boldmath$\nabla$}^2_R+V_{\scriptsize{\rm
conf}}(\textbf{R}).
\end{eqnarray*}
Here $V(r)$ is the screened Coulomb potential, and
$V_{\scriptsize{\rm conf}}(\textbf{R})$ describes the confinement
of the exciton center-of-mass (COM) motion within a sphere of
radius $R_{0}$. With the above Hamiltonian, we are looking for a
solution of the constitutive equation (1) in terms of
eigenfunctions of the Hamiltonians $H_{\scriptsize{\textbf{r}}}$,
and $H_R$. The first mentioned operator is the Hamiltonian
of a hydrogen-like atom, with the eigenfunctions
\begin{equation}
\psi_{n\ell m}(\textbf{r})= N_{n\ell}R_{nl}(r)Y_{\ell
m}(\theta,\phi),
\end{equation} where $Y_{\ell
m}(\theta,\phi)$ are spherical harmonics, and the detailed form of
the eigenfunctions is well-known [4].
For the second Hamiltonian,we will use the no-escape boundary
conditions for the COM motion, so that the eigenfunctions will
have the form
\begin{equation}
\Psi_{NLM}(\textbf{R})=A_{NL}j_{L}(k_{LN}R)Y_{LM}(\Theta,\Phi),
\end{equation}
where $j_L(kR)$ are the spherical Bessel functions, $k$ results
from the equation
\begin{equation}
k_{LN}R_{0}=x_{L,N},\end{equation} where $x_{L,N}$ are the zeros
of $j_L$, and $A_{NL}$ are the normalization factors
$$A_{NL}=\sqrt{\frac{2}{R_{0}^3}}\left[j_{L+1}(k_{LN}R_{0})\right]^{-1}.$$

Assuming the harmonic time dependence of the quantities ${\mathcal
Y},\textbf{E}$, we obtain the constitutive equation in the form

\begin{eqnarray}\label{constCOM}
&&(H_{eh}-\hbar\omega-i{\mit\Gamma}){\mathcal
Y}(\textbf{r},\textbf{R})=\textbf{M}(\textbf{r})\textbf{E}(\textbf{R}).\end{eqnarray}
Inserting the expansion
\begin{equation}
{\mathcal Y}(\textbf{R},\textbf{r})=\sum\limits_{n\ell m
NLM}c_{n\ell m NLM}\psi_{n\ell
m}(\textbf{r})\Psi_{NLM}(\textbf{R})\end{equation} into Eq.
(\ref{constCOM}) we obtain the values of the expansion
coefficients $c$. With the exciton amplitude, we can
determine the mean effective QD susceptibility from the relation
(3), where we have inserted the dipole density
$\textbf{M}(\textbf{r})$ defined in Ref. [3], and assumed
the long wave approximation. The final result is presented in Eq.
(5). 

\section*{References}
[1]	K. P. O'Donnell and X. Chen, Appl. Phys. Lett. \textbf{58}, 2924 (1991).\newline
[2]	T. Kazimierczuk, D. Fr\"{o}hlich, S. Scheel, H. Stolz, and M. Bayer, Nature \textbf{514}, 343 (2014).\newline
[3] Sylwia Zieli\'{n}ska-Raczy\'{n}ska, Gerard Czajkowski, and David Ziemkiewicz, Phys. Rev. B \textbf{93}, 075206 (2016).\newline
[4] L.D. Landau, E.M. Lifshitz,  \emph{Quantum Mechanics, Non Relativistic Theory 3rd ed.} (Pergamon Press, Oxford, 1962).